\documentclass[aps,prd,reprint,groupedaddress,nofootinbib,a4paper, 11pt]{revtex4}
  \usepackage{geometry}
\usepackage{hyperref}
\usepackage{tikz, amsmath} 
\usetikzlibrary{calc}
\usepackage{bbm}
\usepackage{xcolor} 
\usepackage{dsfont}
\usepackage{subfigure}
\usepackage{hyperref}
\usepackage{microtype}
\usepackage{caption}
\usepackage{amsmath} 
\usepackage{float}
\usepackage{graphics}
\usepackage{psfrag}
\usepackage{color}
\usepackage{slashed}
\usepackage{subfigure}
\usepackage{graphicx}
\usepackage{array,multirow}
\usepackage{ulem}
\usepackage{amsfonts}
\usepackage{amssymb}
\usepackage{subfigure}
\usepackage{multirow}
\usepackage{mathtools}

\usepackage{dcolumn}
\usepackage{bm}
\usepackage{slashed}

\usepackage{epsfig}

\newcolumntype{L}[1]{>{\raggedright\let\newline\\\arraybacksslash\hspace{0pt}}m{#1}}
\newcolumntype{C}[1]{>{\centering\let\newline\\\arraybackslash\hspace{0pt}}m{#1}}
\newcolumntype{R}[1]{>{\raggedleft\let\newline\\\arraybackslash\hspace{0pt}}m{#1}}

\newcommand{\beq}{\begin{eqnarray}}
\newcommand{\eeq}{\end{eqnarray}}

\newcommand{\bmp}{\noindent\begin{minipage}{16cm}}
\newcommand{\emp}{\end{minipage}\vskip 7mm} 

\usepackage{dcolumn}
\usepackage{bm}
\usepackage{slashed}

\usepackage{epsfig}

\usepackage{hyperref}

\newcommand{\bea}{\begin{eqnarray}}
\newcommand{\eea}{\end{eqnarray}}

\newcommand{\ba}{\begin{eqnarray}}
\newcommand{\ea}{\end{eqnarray}}

\begin{document}

\title{Exact Diffeomorphism  in Warped Space}

\author{Haiying Cai$^{1}$}
\email{hcai@korea.ac.kr}
\affiliation{
$^1$Department of Physics, Korea University, Seoul 136-713, Korea
}

\begin{abstract} 
In this paper, we study the manifestation of exact diffeomorphism in warped extra dimension, with an emphasis on the Randall-Sundrum model.  Utilizing the covariant  form of  variation  $\delta g_{MN}  = \nabla_M   \xi_N +  \nabla_N   \xi_M$, we derive the nonlinear transformation rules for the metric fields in the unitary gauge.  As an off-shell symmetry, diffeomorphism governs the interaction structure by connecting the neighboring orders of  bulk action expansions. Our analysis unveils  that an on-shell diffeomorphism  exists prior to radion stabilization, but is broken by the Goldberger-Wise  mechanism. This on-shell symmetry  enforces  a mass-related  equation  $F'- A' G =0$ to eliminate one degree of freedom,  while ensuring  the  5D action  invariant up to a surface term.  Therefore,  we  conjecture  that  the on-shell diffeomorphism actually   protects the radion field from acquiring mass.
\end{abstract}

\maketitle

\section{Introduction}

Warped extra dimension models within the context of Anti-de Sitter (AdS) space offer a compelling framework for addressing long-standing puzzles in particle physics, such as the Planck hierarchy problem and the weakness of gravity coupling~\cite{Randall:1999vf, Randall:1999ee, Batell:2008me, Gherghetta:2010he}. A fascinating aspect of these models is the correspondence between the weakly coupled five-dimensional theory and an approximately conformal field theory (CFT) residing on the boundary\cite{Maldacena:1997re, Witten:1998qj, Gubser:1998bc}. From the AdS/CFT perspective, a mass scale should   break the conformal symmetry, which is achieved in the extra dimension through either a hard wall IR cut-off, as in the Randall-Sundrum model~\cite{Randall:1999vf, Randall:1999ee}, or a dynamical IR cut-off, as in the soft-wall model~\cite{Batell:2008me, Gherghetta:2010he}. These models have been  extensively explored in the literature,  particularly due to their profound influence on  our understanding of  gravitational wave,  early universe cosmology and dark matter~\cite{Randall:2006py, Konstandin:2011dr, Baratella:2018pxi}.  In addition to  distinct phenomena, symmetry plays a pivotal role in warped extra dimension models. This paper aims to elucidate the concept of diffeomorphism, i.e. the  invariance originating  from  $x^M \to x^M + \xi^M$ and its consequences in relevant physics. For convenience, we will focus our discussion on the Randall-Sundrum (RS) model with the Goldberger-Wise (GW) mechanism~\cite{Goldberger:1999uk, Goldberger:1999un}, although general conclusions  apply equally to other warped theories such as the soft-wall models.

The  diffeomorphism in extra dimension has been primarily  discussed in  its linearized  approximation~\cite{Pilo:2000et, Gherghetta:2011rr, Dolan:1983aa}, where gauge parameters are used to eliminate redundant degrees of freedom. One specific interpretation relates to the geometric Higgs mechanism for massive gravitons. It has been discovered that the Kaluza-Klein (KK) spectrum exhibits an infinite parameter symmetry, which is spontaneously broken into Poincaré $\otimes$ U(1) by the vacuum configuration of the RS metric $g_{MN}^{(0)}$~\cite{Dolan:1983aa}. Alternatively, this is interpreted in ~\cite{Lim:2007fy, Lim:2008hi, Chivukula:2022kju} as two $\mathcal{N}=2$ supersymmetries hidden in the 4D mass spectrum. Utilizing the linearized 5D diffeomorphism, the spurious KK modes of $g_{\mu 5}$ and  scalar field can be eliminated by the broken gauge parameters, equivalently absorbed by the corresponding spin-2 KK graviton. In this framework, the residual symmetry is reduced to the 4D diffeomorphism observed by the zero mode of the graviton, with the physical radion left invariant. However, this interpretation is limited to the quadratic order expansion of the 5D action.

Analogous to the 4D diffeomorphism for the graviton action~\cite{Hinterbichler:2011tt}, the exact 5D diffeomorphism includes a nonlinear component involving the derivative form of $\xi^M$ multiplied by metric perturbation fields, which is crucial for ensuring that the full action remains invariant. Unlike the linearized  diffeomorphism depicted in~\cite{Dolan:1983aa}, the exact diffeomorphism is observed in an off-shell manner that does not require  to obey the equations of motion and imposes no constraints on the Kaluza-Klein spectrum.  In this paper, we derive the nonlinear diffeomorphism for the metric fields without approximation in the unitary gauge, specifically $g_{\mu 5} = 0$ in the RS model.  As anticipated,  our result shows that the diffeomorphism in the unitary gauge does not mix physical fields with different spins.  Moreover, due to  nonlinearity, this off-shell symmetry connects  the neighboring orders of bulk action expansions via a recurrence relation~\cite{Cai:2023mqn}, effectively governing the interaction structure  in the theory.

The paper is organized as follows. In Section~\ref{sec:invariance}, we present the proof that the 5d action of RS model is invariant under the diffeomorphism without expanding the Lagrangian.  Following that proof, we detail the derivation of the nonlinear transformation rules for the metric fields in Section~\ref{sec:nonlinear}.  We deduce that for  the bulk Lagrangian $\sqrt{g} \mathcal{L}  = \sum_n \mathcal{\hat L}^{(n)} $, the nonlinear variation of  $ \mathcal{\hat L}^{(n)}$ accompanied  with the linear variation of $ \mathcal{\hat L}^{(n+1)}$   results in a total derivative of $\xi^M$ times the lower order expansion. Consequently,  after summing  over all the orders, the diffeomorphism variation of the full action amounts to a surface term. Then  in Section~\ref{sec:radion}, we define an on-shell diffeomorphism by imposing a specific equation of motion constraint and show that the breaking of this on-shell symmetry leads to the radion acquiring its mass. Finally, we present our conclusions in Section~\ref{sec:conclusion}.

\section{Diffeomorphism Invariance of 5d action}~\label{sec:invariance}

We start with a brief review of  the Randall-Sundrum model stabilized by the GW mechanism. The 5d action is written as:
\beq
&&  S =  -   \frac{1}{2 \kappa^2} \int d^5 x  \sqrt{g} \, {\cal R} + 
 \frac{1}{2} \int d^5 x  \sqrt{g}  g^{IJ}  \partial_I \phi \partial_J \phi
\label{Act}    \\ && \quad - \frac{1}{2} \int d^5 x \sqrt{g} V(\phi) + S_{\rm{brane}}  \,, 
 \nonumber \\
&& S_{\rm brane} = -  \int dx^5 \sum_{i} \sqrt{g_4} \lambda_{i}(\phi) \delta(y - y_i) \, \label{Act:brane}
\eeq
where $y$ is the  fifth dimension coordinate  and  the the capital Latin indices $(I, J) \in (\mu, 5)$  span
 all the dimensions, with the Greek index $\mu$  assigned for the Minkowski spacetime.  Eq.(\ref{Act}) consists  of  Einstein-Hilbert action accompanied with the  GW scalar terms. In particular, the brane action  Eq.(\ref{Act:brane}) is required to satisfy the jump condition at the boundary $y_i = \{0, L\}$~\cite{Csaki:2000zn, Cai:2021mrw}.  By adjusting $V(\phi)$ and $\lambda_i(\phi)$ in this set up,   the bulk scalar will develop a $y$-dependent VEV $\phi_0(y)$, which reacts  on the metric such that the radion field acquires its mass~\cite{Goldberger:1999uk, Goldberger:1999un, Csaki:2000zn, Kofman:2004tk}.  The line element in the RS model  is:   
\beq
d s^2  &=&  e^{-2A -2F}  \hat g_{\mu \nu} d x^\mu d x^\nu - \left[ 1+G \right]^2 dy^2  \label{metric} \\
\hat g_{\mu \nu} &=& \eta_{\mu \nu} + h_{\mu \nu}  \nonumber
\eeq
where the  metric  observes  an $S^1/Z_2$ orbifold symmetry and $\eta_{\mu \nu} = \left(+, -,-,-\right)$ is for the Minkowski spacetime. Note that  this is most general parametrization in the unitary gauge which can decouple the graviton from the radion.

Now we are going to  demonstrate that  Eq.(\ref{Act})  is invariant  under  5d diffeomorphism.  In fact,  diffeomorphism involves a pushforward followed by  coordinate transformation back.  Under the  pushforward  $X^I \to X^I + \xi^I(X)$,   an infinitesimal  diffeomorphism variation of tensor field  is generated by Lie derivative.   Since Eq.(\ref{Act}) merely depends on the metric $g_{MN}$ and the GW scalar $\phi$, we will start with the corresponding transformation:
\begin{eqnarray} 
&& \delta g^{MN} = - \left(\nabla^M \xi^N + \nabla^N \xi^M \right)  \label{dg5} \\
&&  \delta \phi = \xi^K \partial_K \phi  \, \label{dphi} 
\end{eqnarray}
where $\nabla^M \equiv g^{MN} \nabla_N$ denotes the covariant derivative and $\xi^M$ is arbitrary in the bulk but vanishes at the boundary. Assuming the bulk action is  $S_{bulk} = \int d^5 x \sqrt{g} \, \mathcal{\hat L} $,  with   $\mathcal{\hat L}$ being a scalar constructed out of tensors,  the  transformation  Eq.(\ref{dg5}-\ref{dphi})  will  lead to  $\delta \left( \sqrt{g} \mathcal{ \hat L} \right) = \partial_M \left( \xi^M \sqrt{g} \mathcal{\hat L} \right) $.   
Firstly, it is easy to prove that the square root of metric determinant  is a total derivative under the infinite diffeomorphism, i.e.
\begin{eqnarray}
\delta  \sqrt{g} &=& -\frac{1}{2} \sqrt{g} g_{MN} \left(  \xi ^K  \partial_K g^{M N}-  g^{N K} \partial_{K} \xi^M - g^{MK} \partial_K\xi^N\right) \nonumber \\
&=&    \xi^M \partial_M \sqrt{g}  +\sqrt{g} \partial_M \xi^M 
= \partial_M \left( \xi^M \sqrt{g} \right)   \label{dg} 
\end{eqnarray}
where  we  apply $\partial_M \sqrt{g} = - \frac{1}{2}   \sqrt{g}   g_{IJ} \partial_M g^{IJ}$ for the first term.  
Then  we need  to prove that  the  transformation  of   Lagrangian density  is  a directional derivative $\delta \mathcal{\hat L} = \xi^M \partial_M \mathcal{\hat L}$ under the 5d diffeomorphism. In the following we explicitly show that  each term in Eq.(\ref{Act})  observes this property.
\begin{itemize}

\item[(1)]
We will start with the variation of  Ricci scalar:
\begin{eqnarray}
\delta \mathcal{R}  = \delta g^{MN}  \mathcal{R}_{MN} +  g^{MN} \delta \mathcal{R}_{MN}  \label{eq:R0} 
\end{eqnarray}
Our strategy is to  recast  Eq.(\ref{eq:R0}) in terms of  covariant derivatives.  Using Eq.(\ref{dg5}), the first term can be expressed as:
\begin{eqnarray}
\delta g^{MN} \mathcal{R}_{MN} =  - \left(\nabla^M \xi^N + \nabla^N \xi^M \right) \mathcal{R}_{MN} \label{eq:R1}
\end{eqnarray}
With lengthy algebra,  the second term in Eq.(\ref{eq:R0}) is transformed to be:
\begin{eqnarray}
g^{MN} \delta \mathcal{R}_{MN} &=& \nabla_M \nabla_N \left( - \delta g^{MN} + g^{MN} g_{IJ} \, \delta g^{IJ} \right) \nonumber \\
&=&  \nabla_M \nabla_N \left(\nabla^M \xi^N + \nabla^N \xi^M - 2 g^{MN} \nabla^K \xi_K \right) \nonumber \\
&=& 2 \nabla^M \left( R_{M N} \xi^N \right)  \label{eq:R2}
\end{eqnarray}
where $ \left(\nabla_M \nabla_N - \nabla_N \nabla_M \right) \xi^M = \mathcal{R}_{M N} \xi^M $ is applied to the last step.
Substituting  Eq.(\ref{eq:R1}-\ref{eq:R2}) into Eq.(\ref{eq:R0}),   we find that:
\begin{eqnarray}
\delta \mathcal{R}  = 2 \, \xi^M \nabla^N \mathcal{R}_{MN} 
\end{eqnarray}
Then we use the contracted Bianchi identity $\nabla^N \mathcal{R}_{MN} = \frac{1}{2} \nabla_M \mathcal{R} $  to perform simplification and this yields:
\begin{eqnarray}
\delta \mathcal{R}  =  \xi^M \nabla_M  \mathcal{R} = \xi^M \partial_M \mathcal{R} \label{dR}
\end{eqnarray}
Note that  since  the 5th dimensional diffeomorphism corresponds to the $\xi^5 \partial_5$ operation,    it is necessary to subtract the  hidden boundary contribution arising from $A''$ terms in the Einstein-Hilbert action. While  the 4d diffeomorphism  works for the  brane terms as well.

\item[(2)]  Using Eq.(\ref{dg5}-\ref{dphi}), the variation of  the scalar kinetic term can be derived  straightforwardly:
\begin{eqnarray}
\delta \left(  g^{MN} \partial_M  \phi  \partial_N  \phi  \right) &=&  2 g^{MN} \partial_M \phi  \, \partial_N \left(\xi^K \partial_K \phi \right) + \xi^K \partial_K g^{MN} \partial_M \phi \partial_N \phi \nonumber \\ 
&-& 2 g^{M K} \partial_K \xi^N \partial_M \phi \partial_N \phi \nonumber \\
&=& \xi^K \partial_K \left( g^{MN} \partial_M \phi \partial_N \phi  \right) \label{dkin}
\end{eqnarray}

\item[(3)]  Because  $\phi$ is a fundamental scalar,  the variation of the GW potential  $V(\phi)$ is simply  a directional derivative:
\begin{eqnarray}
\delta V(\phi) = \frac{\partial V}{\partial \phi}  \delta \phi = \xi^K  \partial_K V(\phi) \label{dV}
\end{eqnarray}

\end{itemize}

Combining Eq.(\ref{dR}-\ref{dV}) with Eq.(\ref{dg}),  one can deduce that $\sqrt g \mathcal{\hat L}$  indeed  transforms as a total derivative under the infinitesimal diffeomorphism. Therefore the variation of 5d action  becomes a surface term:
\beq
\delta \left( \int d^5 x \sqrt g  \mathcal{\hat L} \right)  =  \int d^5 x \partial_M \left( \xi^M \sqrt{g} \mathcal{\hat L}  \right) =0 \, \label{invariance}
\eeq 
that is zero because of $\xi^M =0$ at the volume boundary.

\section{Nonlinear transformation} \label{sec:nonlinear}
When  only the VEVs of the metric and scalar are taken into account  in Eq.(\ref{dg5}-\ref{dphi}), one obtains the linearized diffeomorphism.  However the linear approximation is not valid  beyond the quadratic order of action expansion.  In fact, the exact  diffeomorphism  contains  a  nonlinear part that is crucial to render the full action invariant.  In this section, we will derive the transformation of  metric perturbation field from  the  covariant form of diffeomorphism  in Eq.(\ref{dg5}).  Due to  the fact  $g^{MK} g_{NK} = \delta^M_N$, Eq.(\ref{dg5}) can be rewritten as:
\beq 
\delta g_{MN}  = \nabla_M   \xi_N +  \nabla_N   \xi_M  \label{dg1}
\eeq
with $\xi_M = g_{MN} \xi^N$. For convenience, we  split the gauge parameter into  $\xi^\mu = \hat \xi^\mu $ and $\xi^5 = \epsilon$.  As we demonstrate in the previous section, as long as the variation of metric observes Eq.(\ref{dg1}), the Einstein Hilbert action is invariant under the diffeomorphism transformation. Hence, in order to derive the exact transformation rules for  the perturbation fields  in the unitary gauge, i.e. $g_{\mu 5} =0$, we will not adopt any approximation. Evaluating the right hand of Eq.(\ref{dg1}),  the  $(\mu\nu)$-part gives:
\beq
\nabla_\mu   \xi_\nu  &=&   \partial_\mu \left(g_{\nu \rho} \hat \xi^\rho \right) - \Gamma^{N}_{\mu \nu}  g_{N M}  \xi^M \nonumber \\
&=& g_{\nu \rho} \partial_\mu  \hat \xi^\rho + \frac{1}{2} \left[  \partial_\mu g_{\rho \nu} - \partial_\nu g_{\rho \mu}   \right]  \hat \xi^\rho \nonumber \\
&+&  \frac{1}{2} \ \xi^M \partial_M g_{\mu \nu}        \label{cd1}
\eeq
where $M, N$ are for all the dimensions, while  the Greek index is confined for Minkowski spacetime. Note that although  $g_{\mu 5} =0$ is imposed on Eq.(\ref{cd1}),  the last term still includes a  $\xi^5 = \epsilon$ transformation,  which originates  from the Christoffel connection component. Similarly the fifth dimension $(55)$-part of Eq.(\ref{dg1})  is:
\beq
\nabla_5   \xi_5  &=&   \partial_5 \left(g_{55}  \epsilon \right) - \Gamma^{N}_{55}  g_{N M}  \xi^M \nonumber \\
&=&  g_{55}  \partial_5 \epsilon  +  \frac{1}{2}  \xi^M \partial_M g_{55}  \label{cd2}
\eeq
Substituting Eq.(\ref{cd1}-\ref{cd2}) into Eq.{\ref{dg1}}, one obtains:
\beq
\delta g_{\mu \nu} &=&  g_{\mu \rho} \partial_\nu  \hat \xi^\rho + g_{\nu \rho} \partial_\mu  \hat \xi^\rho + \xi^M \partial_M g_{\mu \nu}   \label{dg4a} \\
\delta g_{55} &=&  2 \, g_{55} \, \partial_5 \epsilon   + \xi^M \partial_M g_{55}  \label{dg5a}
 \eeq
Then the  expansion of left-hand of Eq.(\ref{dg1})  yields:
 \beq
\delta g_{\mu \nu}  &=& e^{-2A- 2F} \delta h_{\mu \nu} - 2 \, e^{-2A -2F} \delta F \hat g_{\mu \nu}  \label{dg4b} \\
\delta g_{55} &=&  -2 (1+G) \, \delta G  \label{dg5b}
\eeq
By comparing Eq.(\ref{dg4a}-\ref{dg5a}) with Eq.(\ref{dg4b}-\ref{dg5b}), we  work out  the transformation rules:
\beq
\delta h_{\mu \nu} &=&   \left(\partial_{\mu} \hat{\xi}_{\nu} + \partial_\nu \hat{\xi}_\mu  \right)  
+ \hat{\xi}^\alpha \partial_\alpha  h_{\mu \nu}   \,  \label{hrule} \\ &+&   \partial_\mu \hat{\xi}^\alpha h_{\alpha \nu} + \partial_\nu \hat{\xi}^\alpha h_{\alpha \mu} + \epsilon   h'_{\mu \nu}    \nonumber  \\
\delta F &=&  A'  \epsilon  +\epsilon F' +\hat{\xi}^\alpha \partial_\alpha F  \, \label{Frule} \\ 
\delta G  &=&  \epsilon' +\partial_5 \left( \epsilon G \right) +\hat{\xi}^\alpha \partial_\alpha G  \,  \label{Grule}
\eeq
with $\hat \xi_\mu = \eta_{\mu \nu} \hat \xi^\nu$ and the prime standing for $\partial_5 = \partial/\partial y$.  in Appendix~\ref{Appendix}, we also dervie the exact diffeomorphism transformation in the conformal coordinate, which confirms that the choice of coordinate does not impact the physics discussed in Section~\ref{sec:radion}.   And  in terms of  $\hat \xi^\mu$ and $\epsilon$,  the transformation for the GW scalar  reads:
\beq
\delta \varphi = \epsilon  \phi'_0  + \epsilon  \varphi' + \hat \xi^\alpha \partial_\alpha \varphi  \label{Vrule} \,.
\eeq
Eq.(\ref{hrule}-\ref{Vrule}) indicates  that the diffeomorphism is a re-parametrization symmetry which do not mix physical fields with different spins. However the gauge parameters are under certain constraints in the unitary gauge.  Since the transformation is necessary to keep the metric in its  original form,  the variation of $g_{\mu 5}$ must  vanish,  which  results in:
\beq
& \delta g_{\mu 5} & \, = \, g_{55} \partial_\mu \epsilon + g_{\mu \nu}  \partial_5 \hat \xi^\nu =0 \nonumber  \\
& \Rightarrow  & \partial_\mu \epsilon =0   \quad  \mbox{and}  \quad  \partial_5 \hat \xi^\nu =0 
\eeq
Thus the unitary gauge diffeomorphism  implies that $\epsilon$ has only $y$-dependence while $\hat \xi^\mu$  depends merely on the Minkowski coordinate.  Note that  the diffeomorphism is observed in an off-shell manner. It is not  appropriate to impose  $\partial^\mu h_{\mu \nu} = h =0$ or $G= 2F$ on Eq.(\ref{hrule}-\ref{Grule}) due to the nonlinearity.

This off-shell symmetry  defines a recurrence  relation  for  the variation of 5d action, by connecting  two neighboring order expansions.  In Eq.(\ref{hrule}-\ref{Vrule}), the first term is the linear approximation, with the remaining terms belonging to the nonlinear part.  Schematically,  one can decompose the variation into  $\delta = \delta_\xi + \delta_\epsilon = \delta^{(1)}  + \delta^{(2)} $, where $\delta_\xi /\delta_{\epsilon}$ contain different gauge parameters and the upper indices $(1)/(2)$ stand for the  linear or nonlinear part respectively.  Let us  expand the bulk action of RS model  in the  following form: 
\beq
S_{\rm bulk} = \int d^5 x \, \sqrt{g}  \mathcal{\hat L} = \int d^5 x  \sum_n  \mathcal{ \hat L}^{(n)} 
\eeq
where  $\mathcal{\hat L}^{(n)} $ is the $n$-th order Lagrangian expansion  that absorbs the metric perturbations from $\sqrt{g}$.  Due to  the orbifold symmetry,  the $A''$ term in $\mathcal{\hat L}$ has to be subtracted out when the fifth dimensional diffeomorphism is invoked. Then for   $\delta \left( \sqrt{g} \mathcal{ \hat L}  \right)= \partial_M \left( \xi^M \sqrt{g} \mathcal{\hat L} \right)$ to hold true,   if and only if   the  Lagrangian expansions  observe a recurrence relation:
\beq  
\delta^{(2)} \mathcal{\hat L}^{(n)} +  \delta^{(1)} \mathcal{\hat L}^{(n+1)}  =  \partial_M \left( \xi^M  \mathcal{\hat L}^{(n)} \right) \label{recurrence}
\eeq
that is a total derivative will be obtained  when one compensates the nonlinear variation of  $n$-th order expansion with the  linear variation of $(n+1)$-th order expansion.  Eq.(\ref{recurrence}) was explicitly verified for the case of $n=2$ in \cite{Cai:2023mqn}. Hence after the summation, we can derive:
\beq
\delta S_{\rm bulk} &=&   \int d^5 x   \partial_M \left( \xi^M \sum_n \mathcal{\hat L}^{(n)} \right)  = 0 
\eeq
that precisely agrees with Eq.(\ref{invariance})  in Section \ref{sec:invariance}. 

\section{Symmetry connection to mass}\label{sec:radion}
In this section, we proceed to investigate how  diffeomorphism invariance constrains  the interaction structure and  the mass generation. Without loss of generality, we first expand the 5d  Lagrangian  given by $\sqrt g  \mathcal{L} = - \frac{\sqrt g  }{2 \kappa^2} \left( \mathcal{R} + 2 \kappa^2 \Lambda \right)$, with $\Lambda = -\frac{6}{\kappa^2}A^{\prime 2}$,  up to the quadratic order.We will focus on  the  potential terms  at the leading orders with two $\partial_5$  derivatives, which  read~\cite{Cai:2021mrw, Cai:2022geu}:
\beq
\mathcal{L}^{(1)}_{tad} &=&   \frac{3 e^{-4A}}{2 \kappa^2}\left(  h  A''   - 8 F  A''   \right) \label{tadpole}  \\  \nonumber \\
\mathcal{L}^{(2)}_m   
&=&-  \frac{e^{-4A}}{2 \kappa^2} \left \{\frac{1}{4} \Big[ h'_{\mu \nu} \,  h'^{\mu \nu} - h'^2 \Big] + \frac{3}{2} \Big[ h_{\mu \nu}  h^{\mu \nu} - \frac{1}{2} h^2 \Big] A''   \right.   \label{mass} \\
&-&  \left. 12  \left(  \left[ F'-  A' G  \right]^2  + 4 F^2 A'' \right) + \mathcal{L}^{(2)}_{\rm mix}  \right\} \nonumber
\\ 
\mathcal{L}^{(2)}_{\rm mix} &=&   3   \left( \left[ F' - A' G \right] h'  + 4 F h A'' \right)   \label{mix}
\eeq
where the $A''$ terms  cancel  the corresponding expansions from the brane action Eq.(\ref{Act:brane}).   The stabilized action Eq.(\ref{Act}) comprises  the same part of Lagrangian,  due to the fact   $V_0 = -\frac{6}{\kappa^2} A'^2 + \frac{1}{2} \phi^{\prime 2}_0$ at the leading order~\cite{DeWolfe:1999cp, Csaki:2000zn}.   Note that tadpoles in Eq.(\ref{tadpole}) vanish in the bulk because of  $A'' = \sum_i  \epsilon \lambda_i \delta(y-y_i)$,  thus the quadratic terms  serve as an initial condition for the bulk recurrence relation in Eq.(\ref{recurrence}).  As a consequence,  the potential terms  at any order are combinations of derivative factors: $(F' - A' G)$,  $A'$, $A''$,  $h'_{\mu \nu}$,  $h'$  and  non-derivative  factors: $F, G$, $h_{\mu \nu}$, $h$. While the $G'$ term  can always be recasted  into this structure using  partial integration,  as verified by  the cubic  expansion  in \cite{Cai:2023mqn}.  In fact, an on-shell diffeomorphism exists prior to Goldberger-Wise  stabilization,   if we impose the massless radion equation $F'-A' G =0$ on the 5d action.   Examining the $F, G$ transformation  in Eq.(\ref{Frule}-\ref{Grule}), we find that:  
\beq
 \delta \left( F'  - A'  G \right) &=&   \partial_5 \left[ \epsilon (F' - A' G )\right]  + \hat \xi^\alpha \partial_\alpha \left( F' - A' G  \right) \nonumber \\ &+&
 A''\epsilon (1+G) \label{constraint}
\eeq
For $F'-A'G =0$ and $\epsilon(y_i) =0 $  at the branes, we obtain $\delta \left( F' - A'  G\right) = 0$  from Eq.(\ref{constraint}). Thus substituting $G = F'/A'$  into the 5d action effectively removes one degree of freedom, while keeping the constrained action  invariant under diffeomorphism. However, the GW mechanism breaks this on-shell diffeomorphism, as the corresponding EOM  is modified to 
 be $F'  - A'  G = \frac{\kappa^2}{3} \phi'_0 \varphi$ for  a massive radion. Using Eq.(\ref{hrule}-\ref{Vrule}), we can derive that:
\beq
&&  \delta \left( F'  - A'  G - \frac{\kappa^2}{3} \phi'_0   \varphi \right) {\bigg \vert}_{ F'  - A'  G = \frac{\kappa^2}{3} \phi'_0 \varphi}  \nonumber \\ &=&  \frac{\kappa^2}{3} \left( \epsilon' \phi'_0 \varphi + \epsilon \phi''_0 \varphi + \epsilon   \phi_0^{\prime 2}  G\right) 
\eeq
which is nonzero because of  $\phi'_0 \neq 0$. This implies if we eliminate $G$ from the 5d action with  that Einstein equation,  the invariance is no long valid. From this perspective, we can conjecture that the breaking of  this on-shell diffeomorphism  causes the radion to acquire  mass.

Analogously, the massless nature of  graviton zero mode is partially own to the 4D diffeomorphism, which guarantees the absence of a gauge-invariant mass term  in the bulk.  However, given that   $\delta^{(2)}_\xi \left(2 h\right) -  \delta^{(1)}_\xi (h_{\mu \nu}  h^{\mu \nu} - \frac{1}{2} h^2) =\partial_\mu \left( \hat \xi^\mu h \right)$,  a brane mass $\propto \left(h_{\mu \nu}  h^{\mu \nu} - \frac{1}{2} h^2\right) A''$ for the graviton in Eq.(\ref{mass}) is actually permitted by the 4D diffeomorphism.  But thanks to the $A''$ cancellation with Eq.(\ref{Act:brane}),  a massless  graviton is  finally secured  in the RS model. Moreover, the GW stabilization maintains this property, as the $A''$ term for the graviton is precisely  removed by the GW potential and brane actions~\cite{Cai:2022geu}.
Note that  KK gravitons with non-constant 5D profiles are not influenced, as their masses originate  from   $ \left(h'_{\mu \nu} \,  h'^{\mu \nu} - h'^2 \right)$, that  is consistent with diffeomorphism.

\section{Conclusion} \label{sec:conclusion}
In this paper, we have presented the  proof of exact diffeomorphism in the unitary gauge of RS  model, highlighting its nonlinearity that ensures the full action invariant up to a surface term. In ~\cite{Dolan:1983aa}, the authors argued that  the radion is massless because it is the Goldstone boson associated with a global scale invariance. As  a novel explanation for this problem as well,  our analysis  reveals the presence of an on-shell symmetry that guarantees the massless nature of radion prior to the GW stabilization. Our results contribute to a better comprehension of the interaction structure in the RS model.

\onecolumngrid
\newpage
\appendix

\section{Diffeomorphism in conformal coordinate}~\label{Appendix} 
The metric in the RS model  can be parametrized in the context of  the conformal coordinate:
\beq
d^2 s = e^{-2A(z)} \left[ e^{-2F} \left(\eta_{\mu \nu} + h_{\mu \nu}\right) dx^\mu dx^\nu - (1+ G)^2 d^2 z  \right]
\eeq
where  $e^{-A} dz = dy$ is employed to transform from the metric in Eq.(\ref{metric}). In the conformal coordinate, Eq.(\ref{dg4a}-\ref{dg5a}) are explicitly written as:
\beq
\delta g_{\mu \nu} 
& = & g_{\mu \rho} \partial_\nu  \hat \xi^\rho + g_{\nu \rho} \partial_\mu  \hat \xi^\rho +  \hat \xi^\rho \partial_\rho g_{\mu \nu} \nonumber \\
& + & \zeta \left[ \partial_z e^{-2A -2F} \left( \eta_{\mu \nu} + h_{\mu \nu} \right) +  e^{-2A-2F}   \partial_z  h_{\mu \nu} \right] \,   \label{dg4c}
\eeq    
and
\beq
\delta g_{55} 
 &=&  - 2 (1+G)^2 e^{-2A} \partial_z \zeta - \hat \xi^\mu \partial_\mu \left[ (1+G)^2 e^{-2A }\right] \nonumber
 \\ &-&  \zeta \partial_z \left[ \left(1+G \right)^2 e^{-2A}\right] \, \label{dg5c}
\eeq
From Eq.(\ref{dg4c}-\ref{dg5c}), one can extract  out the component field transformation rules:
\beq
\delta h_{\mu \nu} &=& \left(\partial_{\mu} \hat{\xi}_{\nu} + \partial_\nu \hat{\xi}_\mu  \right) + \partial_\mu \hat{\xi}^\alpha h_{\alpha \nu} + \partial_\nu \hat{\xi}^\alpha h_{\alpha \mu} \,  \label{dfh}   \\ &+&  \hat{\xi}^\alpha \partial_\alpha  h_{\mu \nu} + \zeta   \partial_z h_{\mu \nu}  \nonumber \\
\delta F &=&   \zeta \partial_z A  + \zeta \partial_z F + \hat \xi^\rho \partial_\rho F \\
\delta G &=&  \partial_z \big [\zeta (1+G)\big]  - \zeta \partial_z A \left(1+ G \right) + \hat \xi^\mu \partial_\mu G \, \label{dfG}
\eeq
where only the variation of $G$ changes due to the coordinate transformation. The fifth dimensional shift is parametrized as $\xi^5 \to  z + \zeta $,  related to the $y$-coordinate in the following way:
\beq
\partial_5 = e^{A} \partial_z \,, \quad \epsilon = e^{-A} \zeta
\eeq
To verify that Eq.(\ref{dfh}-\ref{dfG}) are correct infinitesimal transformation, we  directly calculate the variation of $\sqrt{g}$ 
in the conformal coordinate, which yields:
\beq
\delta_{\zeta} \sqrt{g}  &=& \delta_{\zeta} \Big( e^{-5A-4F} \sqrt{\hat g} (1+G) \Big) \nonumber \\
 &=&   e^{-5A-4F} \Big[ (\delta_{\zeta} G - 4 (1+ G) \delta_{\zeta} F) \sqrt{\hat g} \nonumber \\
 &+& (1+ G) \delta_{\zeta}\sqrt{\hat g} \Big]  \label{dfg}
\eeq
Substituting Eq.(\ref{dfh}-\ref{dfG}) into Eq.(\ref{dfg}),  then we obtains that:
\begin{eqnarray}
\delta_{\zeta} \sqrt{g} 
 &=& e^{-5A-4F} \Big[\partial_z \left(\zeta (1+G)\right)- 5 \partial_z A \zeta (1+G)  \Big] \sqrt{\hat g}  \nonumber \\
&+ &  e^{-5A-4F} (1+G) \zeta \left[  -4  \partial_z F \sqrt{\hat g} +   \partial_z \sqrt{\hat g} \right]  \nonumber \\
&=& \partial_z (\zeta \sqrt{g}) 
\end{eqnarray}
As anticipated,  the transformations of metric fields  precisely recover  Eq.(\ref{dg}). Note that  Eq.(\ref{hrule}-\ref{Grule}) in the $y$-coordinate also pass this simple verification:
\beq
\delta_{\epsilon} \sqrt{g}  &=& \delta_{\epsilon} \Big( e^{-4A-4F} \sqrt{\hat g} (1+G) \Big) \nonumber \\
 &=& e^{-4A-4F} \left[ \partial_5 \left(\epsilon(1+G)\right) - 4 A' \epsilon (1+G)  \right] \sqrt{\hat g}  \nonumber \\
&+& e^{-4A-4F}  (1+G) \epsilon \,\left[  - 4  F' \sqrt{\hat g} + \partial_5 \sqrt{\hat g}  \right]  \nonumber \\
&=& \partial_5 (\epsilon \sqrt{g}) 
\end{eqnarray}
Using the background Einstein equation $\partial_z^2 A + (\partial_z A)^2 =0$ and $(\mu 5)$ Einstein equation $\partial_z  F - G \, \partial_z A  =0  $,  we  find that  the transformation rules  satisfy:
\beq
\delta \left(  \partial_z  F - G \, \partial_z A   \right) =0 \,.
\eeq
Thus  an on-shell diffeomorphism can be defined in the conformal coordinate as well.

\bibliographystyle{JHEP}
\bibliography{warped}

\end{document}